\documentstyle[aps,prb,preprint]{revtex}
\begin{document}
\flushbottom

\title{Disordered Bosons: Condensate and Excitations}
\author{Kanwal G. Singh and Daniel S. Rokhsar}
\address{ Department of Physics, University of California, Berkeley,
California 94720 }
\date{\today}
\maketitle

\begin{abstract}

The disordered Bose Hubbard model is studied numerically within
the Bogoliubov approximation.  First, the spatially varying
condensate wavefunction in the presence of disorder is found by
solving a nonlinear Schrodinger equation.  Using the Bogoliubov
approximation to find the excitations above this condensate, we
calculate the condensate fraction, superfluid density,
and density of states for a two-dimensional disordered system.
These results are compared with experiments done with ${}^4{\rm He}$
adsorbed in porous media.

\end{abstract}

\section{Introduction} 

By definition, superfluids are robust to the introduction of weak
microscopic disorder.  A flowing superfluid is characterized by
a macroscopic wavefunction whose phase varies across the sample;
as long as this condensate wavefunction remains well defined,
disorder cannot lead
to the degradation of currents for topological reasons.\cite{Leggett}
With increasing disorder, however, the rigidity of the superfluid
towards
phase variations is reduced.  For sufficiently large disorder
superfluidity
is eventually destroyed even at zero temperature, resulting in a Bose
insulator (the ``Bose glass'').\cite{Hertz,Thierry,FF,FFGW,Weichman}

Disordered Bose condensates can be realized experimentally by
superfluids\cite{granular} in random media, such as $^4{\rm He}$
films adsorbed on porous Vycor glass.\cite{Reppy,Moses,Fin,Gillis}
The first few monolayers of adsorbed $^4{\rm He}$ are not superfluid,
even at low temperatures, and form an ``inert'' insulating layer of
bosons localized by disorder.
As the coverage is increased, a transition from this Bose insulator
to a superfluid phase is observed.\cite{Reppy}
Crudely speaking, the first few monolayers are comprised of bosons
occupying non-overlapping localized states, which screen the
microscopic
disorder of the porous glass for subsequently added bosons.
Added bosons feel a smoother potential that is the sum of the
initial random potential plus a Hartree repulsion from the localized
particles.
When the disorder is sufficiently well screened, condensation into an
extended state occurs.\cite{inertness}

Of course, this picture is an oversimplification:
the $^4{\rm He}$ atoms in the ``inert'' layer are indistinguishable
from those in the condensate, and the true many-body wavefunctions
must be completely symmetric with respect to particle interchange.
Exchange between the ``inert'' and ``condensed'' bosons can be
important,
especially near the insulator-superfluid transition.
The computational problem with this scheme is that, unlike the case of
fermions, which by the exclusion principle must populate orthogonal
states,
bosons actually prefer to be in non-orthogonal states to optimize
their
effectively attractive exchange interactions.
The need to symmetrize thwarts controlled general Hartree-Fock
calculations.

We present here numerical calculations of the properties of
highly disordered Bose condensates using the
Bogoliubov\cite{Bogoliubov}
approximation, which has been formulated for disordered systems
by Lee and Gunn\cite{LeeGunn} and considered in the weak disorder
limit by Meng and Huang.\cite{MengHuang,badHuang}
Although the Bogoliubov method is strictly valid only in the limit
of weak repulsive interactions, we will consider strongly interacting
systems as well (in the presence of arbitrary disorder) in an attempt
to address the qualitative features of Bose systems in random media
in a quasi-analytic fashion.  Previous theoretical approaches include
numerical simulations,\cite{GGB,Nandini,miloje,erik} scaling analysis,
\cite{FFGW} renormalization group
calculations,\cite{Thierry,RGpaper,MaZ}
and perturbative methods.\cite{LZ}

In the Bogoliubov approximation\cite{LeeGunn} the disordered
potential is
screened by bosons occupying a delocalized condensate wavefunction
which
has larger amplitude where the random potential is deep.
This non-uniform condensate is macroscopically occupied, and
fluctuations
into and out of it are considered due to residual interactions.
These effects deplete the condensate non-uniformly, and lead to
a spectrum
of collective, phonon-like excitations.

The Bogoliubov scenario resembles the
heuristic ``inert layer'' picture discussed above, but is constructed
in the reverse order.
{\it First} the condensate is determined, and {\it then} the (possibly
localized) non-condensate part of the many-body wavefunction is
considered.
This localized, uncondensed part of the ground state corresponds to
the ``inert layer'' discussed above, and can be crudely thought of
as the zero temperature ``normal'' fluid excited from the condensate
by disorder rather than thermal fluctuations (see Sec. VI
for a more precise definition).
The advantage of the Bogoliubov approach is that exchange between the
``condensate'' and the ``normal'' fluid is included naturally.  Its
disadvantage is that interactions within the normal fluid are
essentially ignored.

The outline of this paper is as follows:
In Section II, we introduce the disordered Hubbard
model for bosons and in
 Section III we solve this model in the Hartree
approximation.
In Section IV we review the Bogoliubov approximation for
disordered bosons.
Sections V and VI present calculations
of the depletion of the condensate and the reduction of the
superfluid density due to disorder, respectively.
Section VII reports calculations of the excitation spectrum
and specific heat of the disordered condensate.
Finally, in Section VIII we summarize our results and discuss
experiments.

\section{The boson Hubbard model}

A simple model for disordered interacting bosons is the
Hubbard model for lattice bosons in a random potential:
\begin{equation}
{\cal H} = - t \sum_{\langle i,j \rangle} b^\dagger_i b_j
+ \sum_i V(i) b^\dagger_i b_i
+ {U \over 2} \sum_i b^\dagger_i b^\dagger_i b_i b_i,
\label{Hubbard}
\end{equation}
where $b_i^\dagger$ ($b_i$) creates (destroys) a boson at
lattice site $i$.
The sum on $\langle i,j \rangle$ extends over all nearest-neighbor
pairs of lattice sites, $U$ is the strength of the repulsive on-site
interaction, and $t$ is a hopping matrix element.
The random potential $V(i)$ is uniformly distributed between
$-\Delta$ and $\Delta$.
The total number of bosons is ${\cal N}$, and the number of lattice
sites is ${\cal V}$; the mean density is then
$n \equiv {\cal N}/{\cal V}$.

As a model for the behavior of ${}^4 {\rm He}$ adsorbed in Vycor on
length scales less than the pore size (several hundred
{\AA}ngstr\"{o}ms), each site could
represent a surface location of atomic dimension, connected to
neighboring sites in a two-dimensional network.
We will consider (\ref{Hubbard}) on two-dimensional square lattices
of up to 306 sites, with periodic boundary conditions.
To study disordered Bose condensates on longer length scales, the
sites of model (\ref{Hubbard}) could themselves be used to represent
pores in Vycor, with a three-dimensional connectivity.
Since 300 sites is still quite a small three-dimensional lattice,
we will only report calculations in two dimensions.

\section{The Hartree condensate}

A simple variational ground state for (\ref{Hubbard}) is the
Har\-tree\cite{HFexplain} state
\begin{equation}
\Psi(r_1, r_2, ... ,r_N) = \phi_0(r_1) \phi_0(r_2) ... \phi_0(r_N),
\label{Hartree}
\end{equation}
where all bosons are condensed into the same real, normalized
single-particle wavefunction $\phi_0(i)$.
The many-body state (\ref{Hartree}) is explicitly symmetric under
particle exchange, as befits a Bose state.
For a translationally invariant system, the single-particle state
$\phi_0(i)$ is independent of $i$, and is simply the
zero-momentum state.
In a disordered system, this will no longer be the
case: $\phi_0(i)$ will
adjust to be larger at the minima of the random potential
and smaller at its maxima.

The expectation value of the Hamiltonian (\ref{Hubbard}) in the
variational state (\ref{Hartree}) is
\begin{eqnarray}
\langle \Psi | {\cal H} | \Psi \rangle =
&-& tN\sum_{\langle i,j \rangle} \phi_0(i) \phi_0(j)
+N\sum_i V(i) \phi_0^2(i)
\nonumber \\
&+& {UN^2 \over 2} \sum_i \phi_0^4(i) .
\label{variational}
\end{eqnarray}
To minimize (\ref{variational}) with respect to the (normalized)
single
particle state $\phi_0$ one must solve the discrete nonlinear
Schrodinger equation
\begin{equation}
-t\sum_{j={\rm nn}(i)}\phi_{\lambda}(j)+ \tilde{V}(i)
 \phi_\lambda(i) =
(\mu_0+\epsilon_\lambda) \phi_\lambda(i) ,
\label{nonlinear}
\end{equation}
where the sum over $j={\rm nn}(i)$ extends over the nearest
neighbors $j$
of site $i$.
The effective single-particle potential $\tilde{V}(i)$
is given by
\begin{eqnarray}
\tilde{V}(i) \equiv V(i) + U N |\phi_0(i)|^2 ,
\label{Vscreen}
\end{eqnarray}
where $\phi_0(i)$ is the single particle ground state of
(\ref{nonlinear}).

For convenience, $\mu_0$ in eq. (\ref{nonlinear}) is chosen so that
$\epsilon_0 \equiv 0$, {\it i.e.,} so that the Hartree excitation
energies $\epsilon_\lambda$ are measured with respect to the energy
required to add a particle to the condensate.
The condensate $\phi_0(i)$ and the ${\cal V}-1$ excited
states denoted
by $\phi_\lambda (i)$ together form an orthonormal basis for single
particle states.  For convenience, sums over $\lambda$ will always
implicitly {\it exclude} the condensate.

We solve (\ref{nonlinear}) and (\ref{Vscreen}) iteratively,
as follows.
Beginning with a trial condensate $\phi_0(i)$
(either the zero-momentum state or the exact state for $t=0$),
we compute the corresponding screened potential $\tilde V(i)$.
The resulting single-particle Schrodinger equation is solved
numerically to obtain a new set of single particle eigenstates.
An improved trial condensate is then created by mixing the initial
guess with the lowest energy eigenstate of the screened potential.
This procedure is repeated until (\ref{nonlinear}) and (\ref{Vscreen})
are simultaneously satisfied.
Simple linear interpolation to obtain a new trial condensate
converges very slowly, if at all.
More rapid, consistent convergence was obtained with the
Broyden mixing method commonly used in electronic structure
calculations.\cite{Pickett}
Achieving convergence is the biggest obstacle in our calculation,
particularly for large disorder, and limits the system sizes we can
consider.

The condensate wavefunction $\phi_0(i)$ accumulates at the minima
of the applied random potential, so that the screened potential
$\tilde{V}(i)$ is smoother than $V(i)$, with shallower minima.
These minima of the screened potential are more uniform than those
of the original potential, with approximately the same depth.
Roughly speaking, variations in the condensate conspire to create a
screened potential which resembles the initial random potential, but
with its deepest minima lopped off, as shown in
Fig. \ref{Vscreen-fig}.
As the density $n$ (or the interaction strength $U$) is increased,
the minima of $\tilde{V}(i)$ become shallower and shallower, since
the screening is then more efficient.  No long-range correlations
are introduced in the screening process, as shown in Fig.
\ref{autocorrelation}.

To address the nature of the condensate and Hartree excited states,
we calculate the participation ratio
\begin{equation}
P[\lambda] \equiv {1 \over {\sum_i |\phi_\lambda (i) |^4 }}
\label{P-ratio-def}
\end{equation}
of state $\lambda$, which measures the number of sites at which
$\phi_\lambda(i)$ is appreciable.

The condensate wavefunction $\phi_0(i)$ is {\it always}
extended,\cite{Hertz}
and ``participates'' in a finite fraction of the lattice sites.
The extended nature of the self-consistent ground state of
(\ref{nonlinear}) is demanded by the following argument:
Assume that the state $\phi_0(i)$ were localized.
Then for a non-vanishing
density of bosons, macroscopically occupying this
single-particle state
as in (\ref{Hartree}) would confine a macroscopic number of
interacting
particles to a finite volume
(the localization volume of $\phi_0(i)$).
The interaction energy of the resulting many-body state would
then vary as the square of the total particle number.
In the thermodynamic limit, however, the total energy should
be extensive.
Thus the assumption of a localized $\phi_0(i)$ yields a contradiction,
and the condensate must be extended.

This argument does not preclude a condensate wavefunction $\phi_0(i)$
which is ``lumpy'' -- {\it i.e.,} one which is a
(nodeless) superposition
of well-separated, localized states.  Strictly speaking, such
a lumpy state
is extended and the resulting Hartree state
(\ref{Hartree}) is still a Bose condensate.
We will see below that these lumpy condensates (found for
strong disorder and weak repulsion)
are particularly susceptible to depletion from scattering out of the
condensate, and have substantially reduced superfluid densities.
An alternative, and perhaps better, variational state could be
constructed by instead placing a few particles in each of a
large number of localized states.  As these
localized states overlap (they need not be orthogonal), however, it
becomes difficult to calculate the energy and other properties of the
properly symmetrized state.  Unfortunately, the numerical
tricks which
enable efficient calculation with determinants fail for permanents.

It is well known that all eigenstates of a generic random potential
in one and two dimensions are localized.  How then can the condensate
$\phi_0$ always be extended?  The loophole that permits this is that
the screened potential $\tilde{V}(i)$ is {\it not} generic, but
has been
tailored to the problem at hand specifically to produce an extended
ground state.  The condensate is not a ``typical'' state,
but one whose peaks and valleys have been fed back into the disordered
potential itself via (\ref{Vscreen}).  The extended nature of the
condensate does not violate any accepted lore of localization.

An analysis of the participation ratio for (two-dimensional) systems
ranging from ${\cal V}=72$ to ${\cal V}=306$
suggests that for small disorder, all states are extended
({\it i.e.}, have a localization length larger than our
largest system).
Fig. \ref{Plambda-2D} shows that the
participation ratio scales with the size of the system.
For sufficiently strong disorder, we find that the participation
ratios of the Hartree excited states become independent
of system size, indicating
that they have all become localized.  Only the condensate
remains extended.
It is interesting that the use of the self-consistent potential
$\tilde{V}(i)$ converts the state of lowest energy (which in a
typical single-particle localization problem would be the {\it most}
localized state) to the unique extended state.

\section{The Bogoliubov approximation}

Given the self-consistent Hartree condensate, $\phi_0(i)$, we can
proceed with the Bogoliubov approximation.
Following Lee and Gunn,\cite{LeeGunn} we expand the boson
field operator
$b_i$ in the complete set of operators $b_0$ and $\{b_\lambda\}$:
\begin{equation}
b_i = \phi_0(i) b_0 + \sum_{\lambda} \phi_{\lambda}(i) b_\lambda .
\label{expand}
\end{equation}
Although interactions and the disordered potential will
both deplete the condensate, in the Bogoliubov approximation this
depletion is assumed to
be small enough that the single-particle state $\phi_0(i)$ is still
occupied by a macroscopic number ${\cal N}_0$ bosons.
To order $1/{\cal N}_0$ we can then replace the creation
and annihilation
operators $b_0^\dagger$ and $b_0$ for this state by
$\sqrt{{\cal N}_0}$.
The total number of bosons in the system is the sum of those in the
condensate and those not in the condensate:
\begin{equation}
{\cal N} = {\cal N}_0 + \sum_{\lambda} b_\lambda^\dagger b_\lambda .
\label{N0def}
\end{equation}
Expanding to first order in the depletion of the condensate,
we then find
\begin{eqnarray}
b_0 &=& \sqrt{{\cal N}_0} = \sqrt{{\cal N}} -
{1 \over {2\sqrt{\cal N}}} \sum_{\lambda} b_\lambda^\dagger
b_\lambda + ...
\label{sqrtN0}
\end{eqnarray}

Inserting (\ref{expand}) and (\ref{sqrtN0}) into the disordered
Hubbard model
(\ref{Hubbard}), and retaining all terms second order in
$b^\dagger _\lambda$
and $b_\lambda$, yields\cite{LeeGunn}
\begin{eqnarray}
{\cal H}_B &=& {\cal N}\mu_0 - {{U{\cal N}^2} \over 2}
\sum_i |\phi_0(i)|^4 +
\sum_{\lambda\lambda'} \epsilon_\lambda b_{\lambda}^\dagger
b_{\lambda}
\label{bogo}
\\
&+& {{U{\cal N}} \over 2} \sum_{\lambda\lambda'} S_{\lambda\lambda'}
( b_\lambda^\dagger b_{\lambda'} + b_{\lambda'}^\dagger b_\lambda +
  b_\lambda^\dagger b_{\lambda'}^\dagger + b_\lambda b_{\lambda'}).
\nonumber
\end{eqnarray}
The first line of (\ref{bogo}) specifies the single-particle and
self-interaction energies of the condensate, and the energy
$\epsilon_{\lambda}$ for adding a particle in the excited
state $\lambda.$
The second line involves the inner product
\begin{equation}
S_{\lambda\lambda'} \equiv \sum_i
\phi_\lambda(i) |\phi_0(i)|^2 \phi_{\lambda'}(i)
\label{overlap}
\end{equation}
of states $\lambda$ and $\lambda'$ weighted by the
condensate density,
which gives the amplitude for (a) single-particle scattering by the
condensate and (b) pair scattering into and out of the condensate.
Since the condensate is non-uniform, these scattering processes
will generally not conserve momentum.

To arrive at (\ref{bogo}) terms cubic and higher order in
field operators
$b_\lambda$ and $b_\lambda^\dagger$ have been discarded,
This is equivalent to the random phase approximation, and includes
interactions between the non-condensate bosons
and the condensate while neglecting interactions among the
uncondensed bosons.
These approximations are
controlled in the dilute, weakly interacting limit in which the
condensate fraction ${\cal N}_0/{\cal N}$ is close to unity (see
Sec. V).
Here we will push the Bogoliubov approximation to its limits,
and hope
that the qualitative results are representative of disordered Bose
condensates.

The quadratic Hamiltonian (\ref{bogo}) can be diagonalized by
canonical transformation to a set of quasiparticle creation and
annihilation operators $\gamma^\dagger$ and $\gamma$ such that
\begin{equation}
[{\cal H}_{\rm B}, \gamma_\mu^\dagger]  = \omega_\mu
\gamma_\mu^\dagger ,
\label{gamma-excitation}
\end{equation}
where $\omega_\mu$ is the quasiparticle excitation energy.
This transformation is accomplished by taking linear combinations of
creation and annihilation operators:
\begin{equation}
\gamma_\mu^\dagger
=\sum_{\lambda} (u_{\mu\lambda} b^\dagger_\lambda +
v_{\mu\lambda} b_\lambda) .
\label{gamma-def}
\end{equation}
Like the index $\lambda$ labeling the Hartree states, the index $\mu$
labeling quasiparticle states runs from 1 to ${\cal V}-1$.

To satisfy (\ref{gamma-excitation}), the
coefficients $u_{\mu\lambda}$,
$v_{\mu\lambda}$ must obey the generalized eigenvalue equation
\begin{eqnarray}
\left( \begin{array}{rr}
A_{\lambda\lambda'} & -B_{\lambda\lambda'} \\
-B_{\lambda\lambda'} & A_{\lambda\lambda'}
\end{array} \right)
\left( \begin{array}{rr}
u_{\mu\lambda'}  \\ v_{\mu\lambda'}
\end{array} \right) \mbox{\qquad \qquad} \nonumber \\
\mbox{\qquad} = \omega_\mu
\left( \begin{array}{cc}
\delta_{\lambda\lambda'} & 0   \\
0 & -\delta_{\lambda\lambda'}
\end{array} \right)
\left( \begin{array}{rr}
u_{\mu\lambda'}  \\ v_{\mu\lambda'}
\end{array} \right),
\label{4nby4n}
\end{eqnarray}
where

\begin{eqnarray}
A_{\lambda\lambda'}&\equiv&\epsilon_\lambda \delta_{\lambda\lambda'}
+ U{\cal N}S_{\lambda\lambda'}, \nonumber \\
B_{\lambda\lambda'}&\equiv&U{\cal N}S_{\lambda\lambda'}.
\label{defs}
\end{eqnarray}
(Summation over the repeated index $\lambda'$ is implied.)
Note that if $({u \atop v})$ is a solution with excitation
energy $\omega$
(corresponding to $\gamma^\dagger$), then $({v \atop u})$
is a solution
with $-\omega$ (corresponding to $\gamma$). The orthonormality
conditions

\begin{equation}
\left( \begin{array}{cc}
u_{\mu\lambda}  & v_{\mu\lambda}
\end{array} \right)
\left( \begin{array} {cc}
\delta_{\lambda\lambda'} & 0  \\
0 & -\delta_{\lambda\lambda'}
\end{array} \right)
\left( \begin{array}{rr}
u_{\mu'\lambda'}  \\  v_{\mu'\lambda'}
\end{array} \right)
= \delta_{\mu\mu'}
\end{equation}
are {\it automatically} satisfied by normalized solutions of
(\ref{4nby4n}), and guarantee that the quasiparticle operators
$\gamma_\mu^\dagger$ obey Bose commutation relations:
$[\gamma_\mu, \gamma_{\mu'}^\dagger] = \delta_{\mu\mu'}$ and
$[\gamma_\mu, \gamma_{\mu'}] = 0$.

The ground state energy $E_G$ in the Bogoliubov approximation is
\begin{eqnarray}
E_G &=& N \mu_0 - {{UN^2} \over {2}} \sum |\phi_0(i)|^4
\label{GSenergy}
\\
&+&
\sum_{\lambda \lambda ' \mu} \Bigl[ (\epsilon_{\lambda \lambda'}
+UNS_{\lambda \lambda'}) v_{\mu\lambda} v_{\mu\lambda'} - UNS_{\lambda
\lambda'}u_{\mu\lambda} v_{\mu\lambda'} \Bigr] .
\nonumber
\end{eqnarray}
The last line gives the zero-point contribution of the quasiparticle
modes.

\section{The condensate fraction}

The ground state wavefunction $|G\rangle$ in the
Bogoliubov approximation
is the state annihilated by all of the quasiparticle
destruction operators
$\gamma_\mu$:
\begin{equation}
|G\rangle =
(b_0^\dagger)^{{\cal N}_0} \Pi_{\lambda\lambda^\prime}
\exp[-M_{\lambda\lambda^\prime}b_\lambda^\dagger
b_{\lambda^\prime}^\dagger ]
|{\rm  vac} \rangle,
\label{ground-state}
\end{equation}
where $|{\rm vac}\rangle$ is the state with no bosons and
$M_{\lambda\lambda'}$
is defined implicitly by
\begin{equation}
\sum_\lambda u_{\mu\lambda} M_{\lambda\lambda'} = v_{\mu\lambda'}.
\end{equation}
The number of
particles in the condensate, ${\cal N}_0$, is determined by
calculating
the mean number of bosons {\it not} in the condensate ($\sum_\lambda
b_\lambda^\dagger b_\lambda$) and subtracting it from the
total particle number (see eq. (\ref{N0def})).

In a translationally invariant system, the condensate
fraction measures
the occupation
of the zero-momentum state.  In a disordered system,
the proper definition
of the condensate fraction is the largest eigenvalue of the
one particle
density matrix.\cite{ODLRO}  The condensate density is then
given by the square
of the off-diagonal long-range order parameter:
\begin{eqnarray}
\lim_{|i-j|\rightarrow\infty}
\langle G | b_i^\dagger b_j | G \rangle
&=& \langle  G |b_i^\dagger | G \rangle
\langle G| b_j | G \rangle \nonumber \\
&=&{\cal N}_0 \phi_0(i) \phi_0(j).
\label{ODLRO}
\end{eqnarray}

Fig. \ref{depletion} shows the condensate fraction,
${\cal N}_0/{\cal N}$,
as a function of disorder for several values of the
interaction strength.
The calculation is done by averaging over 7 realizations of disorder
on $L_x \times L_y$ lattices where $L$ ranges from
$8$ to $18$.  We then extrapolate to the thermodynamic limit.
Even in the absence of disorder, particles are scattered out of the
condensate as a result of their mutual interactions, and
${\cal N}_0/{\cal N}$ is less than unity.
For weak disorder, the number of bosons in the condensate stays
roughly
fixed, while the condensate wavefunction itself is distorted to
accommodate the random potential.

This insensitivity of the condensate fraction to weak disorder is a
crude criterion for superfluidity, although the proper quantity to
consider is the superfluid density (see below).
Fig. \ref{depletion} shows that as the interaction strength $Un/t$
increases, the system becomes more robust to the addition of disorder,
so that larger values of $\Delta$ are required to further deplete the
condensate beyond the effect of interactions alone.

At large values
of disorder, the condensate fraction drops to zero.  As the condensate
fraction becomes small, the approximation of truncating the Bogoliubov
Hamiltonian (\ref{bogo}) at quadratic order becomes worse and worse,
and our calculations cannot be considered quantitative.  Nevertheless,
our calculation suggests that for sufficiently high disorder the
condensate is destroyed, and a ``Bose glass'' is reached.
Although the logic leading to it breaks down for ${\cal N}_0 = 0$,
the Bogoliubov ground state (\ref{ground-state}) with
vanishing condensate
density is a potentially useful variational state for the Bose glass.

In the Hartree calculation of Sec. II, the
applied random potential $V(i)$ is screened by
${\cal N} U |\phi_0(i)|^2$,
which is equivalent to assuming that all of the bosons are in the
condensate.  As the condensate is depleted, does this estimate of the
screened potential continue to hold?  To check this, we compare
the ground state expectation value of the boson density at site $i$,
$\langle G | b_i^\dagger b_i | G \rangle$, with the density obtained
in the Hartree approximation, ${\cal N} |\phi_0(i)|^2$.
As Fig.  \ref{densities} shows, the density in the Bogoliubov
approximation faithfully tracks the density in the Hartree
approximation.
Even when the condensate is significantly depleted, the particles
scattered from it remain in their original vicinity, and continue to
screen the initial random potential as if they had remained in the
condensate.

\section{Superfluid density}

The superfluid density of a Bose condensate distinguishes between
the low-frequency, long-wavelength transverse and longitudinal
responses of the system.
(This quantity should not be confused with the condensate fraction
discussed above, which is a ground state expectation
value that measures the degree of off-diagonal long-range order.)
A longitudinal probe corresponds to boosting the system, and
the entire fluid responds.
A low-frequency transverse probe corresponds to a slow rotation
of the system, which only couples to the normal fluid, leaving
the superfluid untouched.
The superfluid density is defined simply as the difference
between the longitudinal and transverse response.
In principle, a Bose system can be superfluid without possessing
true off-diagonal long-range order, the canonical example being the
two-dimensional Bose liquid at non-zero temperature below the
Kosterlitz-Thouless transition, which has only algebraic correlations.

The zero-temperature, zero-frequency, current-current response tensor
$\chi_{ij}({\bf q},\omega=0)$ is given by the Kubo formula\cite{PN}
\begin{equation}
\chi_{ij}({\bf q},\omega =0) \equiv -2 \sum_m
{{ \langle G | J_j({\bf q})| m \rangle \langle m |
J_i({\bf q}) | G \rangle}
\over {\omega_m}} .
\label{def-chi}
\end{equation}
The sum extends over all intermediate excited states $m$,
and the lattice current operator ${\bf J}({\bf q})$ is defined by
\begin{equation}
J_i({\bf q}) = 2t \sum_{\bf k} \sin(k_i+{q_i \over 2})
b_{\bf k}^\dagger b_{\bf{k}+\bf{q}} ,
\label{current}
\end{equation}
where $ b_{\bf k} = \sum_i e^{i{\bf k}\cdot{\bf r}_i} b_i $.
In principle the direct evaluation of $\chi_{ij}$ is straightforward
given a complete knowledge of the excited states $|m\rangle$.

In the continuum, the longitudinal response in the long
wavelength limit
is required by the the $f$-sum rule to satisfy
\begin{equation}
\lim_{q\to 0} \chi_{xx}(q\hat{\bf x}) =
-2t{\cal N} ,
\label{longitudinal}
\end{equation}
{\it i.e.,} the entire fluid participates in longitudinal flow.
On a lattice, the $f$-sum rule is modified\cite{fsum} so that
\begin{equation}
\lim_{q\to\infty} \chi_{xx}(q\hat{\bf x}) =
- 2t {\cal N}_{\rm eff} = -2t[{\cal N} -\sum_{\bf k}
{\epsilon_{\bf k}
\over {4t}} \langle n_{\bf k} \rangle],
\label{longitudinal2}
\end{equation}
where $\epsilon_{\bf k}$ is the tight-binding dispersion given
below in (\ref{epsilonk}).  (This sum rule holds in the presence of
arbitrary disorder.) Eq.
(\ref{longitudinal2}) implicitly defines ${\cal N}_{\rm eff}$.
Even though ${\cal N}_{\rm eff} \neq {\cal N}$, the longitudinal
response of a lattice system still corresponds to the entire fluid.

The transverse response of a Bose liquid is only due to
the ``normal fluid,''
since the superfluid component of the system can only participate in
irrotational ({\it i.e.,} longitudinal) flow.  Thus we
can define the number of bosons in the normal fluid, ${\cal N}_n$, by
\begin{equation}
\lim_{q\to 0} \chi_{xx}(q\hat{\bf y}) \equiv -2t{\cal N}_n.
\label{transverse}
\end{equation}
This definition also holds in the continuum, with $t$ replaced
by $\hbar^2/2m$.

In a non-superfluid system, the long-wavelength longitudinal and
transverse responses are identical.   Superfluidity occurs when the
two responses become different.  The superfluid (number) density
$n_s$ is then defined by the difference between the longitudinal and
transverse response, per unit volume:
\begin{equation}
n_s = { {\cal N}_s \over {\cal V}}
\equiv
{{ {\cal N}_{\rm eff} - {\cal N}_n } \over {\cal V}}.
\label{ns-def}
\end{equation}

The longitudinal and transverse response functions
$\chi_{xx}(q{\hat{\bf x}})$ and $\chi_{xx}(q{\hat{\bf y}})$
can be easily calculated numerically in the Bogoliubov approximation.
The excited states $|m\rangle$ entering
(\ref{def-chi}) are then all one and two quasiparticle states.
Because the Bogoliubov approximation does not conserve
particle number,
the lattice $f$-sum rule (\ref{longitudinal2}) is not satisfied.
Explicit evaluation of (\ref{def-chi}) shows that
$\lim_{q\to\infty} \chi_{xx}(q\hat{\bf x})$ tends to the number of
bosons in the condensate, ${\cal N}_0$, rather than the
total particle number.\cite{lattice}
(Note that ${\cal N}_0$ is {\it not} the same as
${\cal N}_{\rm eff}$ in
(\ref{longitudinal2}).)

The normal fluid density, obtained by explicit calculation in the
Bogoliubov approximation of the transverse response
function $\chi_{xx}(q{\hat{\bf y}})$, appears to be more trustworthy.
${\cal N}_n$ vanishes in the translationally invariant case
($V(i)=0$), as
expected.  We therefore follow ref. \onlinecite{MengHuang}
and adopt (\ref{ns-def})
as our operational definition of the superfluid density in the
Bogoliubov approximation.

Fig. \ref{chi} shows the transverse response function
for a $12 \times 13$
lattice averaged over $6$ realizations of various weak disorder.
Such small systems were used because the evaluation of the response
tensor $\chi_{ij}({\bf q},\omega=0)$ in the presence of
disorder requires
four nested sums over quasiparticle states for each {\bf q} and is
computationally very expensive.
Fig. \ref{Neff} shows ${\cal N}_{\rm eff}$ and ${\cal N}_n$ for
the same systems.
${\cal N}_{\rm eff}$ is calculated by explicitly evaluating the
right hand side of
(\ref{longitudinal2}), whereas ${\cal N}_n$ is obtained
from extrapolating
the transverse response from Fig. \ref{chi} to $q=0$.
The difference between
${\cal N}_{\rm eff}$ and ${\cal N}_n$ is ${\cal N}_s$.
For weak disorder ($\Delta/t < 5$), the fluctuations from realization
to realization are small.  With increasing disorder, however, these
fluctuations become quite large, as seen by the error bars in
Fig. \ref{chi} which represent sample-to-sample fluctuations.
Note that the zero-frequency transverse response shows little
dependence on momentum in this approximation.

An alternative definition of superfluid density is as a stiffness to
variations in the phase of the condensate wavefunction.\cite{Twist}
Such a phase variation imposes a superfluid velocity on the system,
\begin{equation}
v_s={\hbar \over m} {\nabla \theta} ,
\end{equation}
where $\theta$ is the phase of the condensate wavefunction.
The total energy of the system increases due to the kinetic energy of
the superflow, and is directly proportional to the density
of superfluid.
The superfluid density is then defined by
\begin{equation}
{{\Delta E} \over {\cal V}} = {{\hbar^2\rho_s} \over {2m^2}}
(\nabla \theta)^2 .
\label{SFtwist-continuum}
\end{equation}

It is easy to estimate the superfluid density in this manner using
the energy of the Hartree state (\ref{Hartree}).  (In principle,
one should also include the change in the zero-point energy of the
quasiparticles  (\ref{GSenergy}), but this calculation
is also costly.)
To impose a twist in boundary conditions, we should change the hopping
matrix elements $t_{ij}$ to $t_{ij}e^{iA_{ij}}$, where $A_{ij}$ is a
vector potential whose sum along a path spanning the sample
is $\theta$.
We should then solve the corresponding new non-linear
Schrodinger equation, and compare the resulting condensate
energies.  Unfortunately, for $\theta \neq {\rm n}\pi$ this requires
solving a complex non-linear Schrodinger equation.

For $\theta=\pi$, the Schrodinger equation (\ref{nonlinear})
remains real,
but a new problem arises.  Consider first the uniform case with
$V(i)=0.$  With a $\pi$ phase twist, the ground state manifold
of (\ref{nonlinear}) is doubly degenerate, and is spanned by
the uniformly
left- and right-moving condensates. This degeneracy frustrates
our iterative convergence scheme, since linear combinations of these
two degenerate solutions have spatially varying densities, driving
even the Broyden method away from convergence.  This problem persists
in the disordered case.

To avoid these complications, we note that for a phase difference
of $\theta = 2\pi$ the Schrodinger equation is unchanged.  In the
course of increasing the phase difference from $0$ to $2\pi$, the
ground state is deformed into the first excited state.  Thus we have
taken the energy $\epsilon_{\lambda=1}$ of the first Hartree excited
state in the absence of a twist to be the energy for introducing
a $2\pi$ phase twist across the sample.
In our problem, the superfluid density can then be computed by
\begin{equation}
{ {{\cal N} \epsilon_1} \over {\cal V}} = 2\rho_s t^2
\big( {{2\pi} \over L} \big)^2.
\label{SFtwist-lattice}
\end{equation}
(Strictly speaking, ${\cal N}_{\rm eff}$ should be used in
place of ${\cal N}$ in (\ref{SFtwist-lattice}), since without
disorder the normal fluid density vanishes, while the longitudinal
response is given by ${\cal N}_{\rm eff}$.  This correction is
comparable in magnitude to the alteration of the zero-point motion
of the quasiparticle energy, which we have also neglected
in obtaining (\ref{SFtwist-lattice}).)

Fig. \ref{twist} shows the superfluid fraction obtained using
(\ref{SFtwist-lattice}) and again extrapolating to the
thermodynamic limit
by averaging over 7 realizations for system sizes
from $L=8$ to $18$.
For weak disorder, when the direct evaluation of the response
function permits reliable extrapolation to ${\bf q}=0$, the two
calculations agree.
As the disorder grows, however, the fluctuations in the response
function (\ref{def-chi}) from sample to sample increase, and beyond
$\Delta/t \sim 10$ the direct calculation of $\chi_{ij}$ is no
longer feasible because of the large number of
samples required to obtain a reasonable statistical average.

As shown in Fig. \ref{twist}, systems with larger interaction
are more robust
to the addition of disorder and thus require
larger $\Delta$ to reduce the
superfluid density.  (This was also the case with the condensate
fraction; compare with Fig. \ref{depletion}.)
In fact, the superfluid fraction
remains substantial even when the Bogoliubov approximation
begins to break
down, {\it i.e.}, when the condensate fraction becomes small.
Note that the superfluid response involves both the bosons in the
condensate and those that have been scattered out of it by
interactions --
when the condensate is accelerated, some of these scattered bosons
accompany it.  In the absence of disorder (and neglecting lattice
effects\cite{lattice}) all bosons participate, and $\rho_s = \rho,$
even though the condensate can be substantially depleted.
As $Un/t$ is increased for fixed disorder, the condensate fraction
${\cal N}_0 / {\cal N}$ is reduced (because of increased scattering
out of the condensate),  while the superfluid fraction $\rho_s / \rho$
is increased (because of decreased sensitivity to disorder when
interactions are strong).

\section{Excitation spectrum}

At long wavelengths, the collective excitations of a uniform Bose
condensate are phonons with a linear dispersion, $\omega = c k.$
For wavelengths comparable to the interparticle spacing, strongly
interacting Bose fluids exhibit a roton minimum, and for even shorter
wavelengths the collective excitations become ill defined,
merging with the multiparticle continuum.\cite{PN}

In the Bogoliubov approximation for lattice bosons without disorder,
the quasiparticle spectrum can be solved analytically.
The tight-binding dispersion is
\begin{equation}
\epsilon_{\bf k} = -2t [cos(k_x) + cos(k_y) - 2] ,
\label{epsilonk}
\end{equation}
and quasiparticle dispersion is
\begin{equation}
\omega_{\bf k} = \sqrt{2Un\epsilon_{\bf k} + \epsilon_{\bf k}^2} .
\label{omegak}
\end{equation}
A linear phonon dispersion holds for wavelengths that are long
enough that (a) $\epsilon_{\bf k}$ is much less than $t$, so that the
tight-binding dispersion is nearly quadratic
($\epsilon_{\bf k} \approx tk^2$), and
(b) $\epsilon_{\bf k}$ is much less than $2Un$, so that the first
term in the square-root in (\ref{omegak}) dominates.
The speed of sound $c$ is then $\sqrt{2Unt}$.

The Bogoliubov approximation is too crude to capture the
roton minimum
found in real strongly-interacting Bose fluids,
and for higher momenta
the excitation energy (\ref{omegak}) rises monotonically.
At short wavelengths pair scattering can be neglected, and the
quasi-particles behave as free particles with a Hartree energy
$\omega_{HF} = tk^2+Un$.

When disorder is introduced, translational invariance is destroyed,
and momentum is no longer a good quantum number.\cite{badHuang}
Are the excitations created by $\gamma_\mu^\dagger$ localized?
Even if the Hartree states $\phi_\lambda$ are localized, the
quasiparticle states created by $\gamma^\dagger_\mu$ need not be --
the condensate is extended, and can mediate non-local scattering via
the inner product $S_{\lambda\lambda'}$ in the Bogoliubov Hamiltonian
(\ref{bogo}).

Since the quasi-particle operators $\gamma_\mu^\dagger$ do not simply
add a boson, but superpose a particle and a ``hole'' (a particle
supplied by the condensate), the participation ratio used for the
Hartree excitations (eq. (\ref{P-ratio-def})) is inappropriate.
Transforming (\ref{gamma-def}) to the site basis, the quasiparticles
are created by
\begin{equation}
\gamma_\mu^\dagger = \sum_{i} (U_{\mu i} b^\dagger_i +V_{\mu i} b_i).
\label{gamma-def-2}
\end{equation}
Adding an excitation in state $\mu$ to the ground state,
there is amplitude $U_{\mu i}$ to create a particle at site $i$ and
$V_{\mu i}$ to create a ``hole'' there.
The net particle density at site $i$ in the one quasiparticle state
$\gamma_\mu^\dagger|G\rangle$ differs from the density of the
ground state itself by
\begin{eqnarray}
\delta n_{\mu i} &\equiv&
\langle G | \gamma_\mu n_i \gamma_\mu^\dagger|G\rangle -
\langle G | n_i |G\rangle
\nonumber
\\
&=&
|U_{\mu i}|^2 + |V_{\mu i}|^2 .
\end{eqnarray}
The corresponding ``participation ratio'' specifying the degree
of delocalization of this density fluctuation is then
\begin{equation}
P[\mu] =
{{(\sum_i \delta n_{\mu i})^2}
\over
{\sum_i \delta n_{\mu i}^2}}.
\label{Pmu-def}
\end{equation}

For an extended excitation, the participation ratio (\ref{Pmu-def})
should scale linearly with the volume of the system; for a localized
excitation, the participation ratio should become independent of the
volume for systems larger than the localization length.
Unfortunately, we could not perform a reliable scaling analysis with
the small systems available to us, and we therefore could not infer
the nature of the excitations created by $\gamma_\mu^\dagger .$
On general grounds, however, we expect the nature of the excitations
in a disordered Bose condensate to be given by the localization
problem
for phonons.\cite{John} Thus in two dimensions, all excitations
should be localized with a
frequency dependent localization length
$\xi(\omega)\sim \exp[A/\omega^2]$
for arbitrary disorder.
(In three dimensions, a mobility edge separates the extended
low energy
phonons from higher energy localized modes.)

Within the Bogoliubov approximation (\ref{bogo}), the
excited states of
the system are independent bosons created by $\gamma_\mu^\dagger$.
If we assume that the temperature is low enough that thermal
fluctuations do not change the excitation spectrum appreciably, but
merely excite the
quasiparticle states according to the Bose-Einstein distribution,
the specific heat is then completely determined by the density of
quasiparticle states per unit energy.
For a density of states $g(\omega)$, the specific heat is
\begin{equation}
{{C(T)} \over k_B} =
\int_0^\infty
{{\omega^2} \over {(k_B T)^2}}
{{g(\omega)e^{\omega/k_BT} \, d\omega}
\over
{[e^{\omega/k_BT} - 1]^2}} .
\label{specific}
\end{equation}

To infer the behavior of the specific heat at low temperatures,
it is useful to introduce the ``integrated density of states''
\begin{equation}
N(\omega) \equiv \int_0^\omega g(\omega') d\omega'
= \sum_\mu \Theta(\omega - \omega_\mu ),
\label{integrated}
\end{equation}
where $\Theta(\omega)$ is the Heaviside step function.
$N(\omega)$ gives the number of states with energy less than
or equal to $\omega$.   As a monotonic function of $\omega$,
$N(\omega)$ is easier to fit than the spiky $g(\omega)$ for
finite systems.
If $N(\omega) \sim \omega^x$, then $C(T) \sim T^x$ for
low temperatures.
For a linear phonon dispersion $\omega = ck$ in a $d$-dimensional box
of linear dimension $L$, the integrated density of states varies as
$N(\omega) \sim (L \omega/c)^d $, so that the specific heat of a
uniform Bose condensate varies as $T^d$ at low temperatures.

For the finite $L_x \times L_y$ lattices we consider, the momenta
${\bf k}$ are restricted to a discrete set of allowed values.
Only a limited number of these values satisfy the condition that
$\epsilon_{\bf k}$ be much smaller than both $ t$ and $2Un$
(or equivalently, that $\omega$ be much smaller than
both $\sqrt{2Unt}$
and $2Un$)
needed for (\ref{epsilonk}) and (\ref{omegak}) to yield the correct
linear dispersion in the absence of disorder.
To obtain enough states to permit a fit to the density of states in
this regime, we are forced to work with large $Un$ even though the
Bogoliubov approximation is uncontrolled in this limit.
The condition that $\epsilon_{\bf k}$ is much less than $t$
guarantees that we avoid the van Hove singularity at the center of the
tight-binding band.  (The van Hove singularity can also be
pushed to higher
energy by the judicious addition of further range hopping
matrix elements
which cancel the $k^4$ terms in (\ref{epsilonk}) and prolong the
$k^2$ dependence of $\epsilon_{\bf k}$.)

In the presence of disorder, the low energy integrated
density of states will
deviate from its pure $\omega^d$ form.  The integrated density
of states
divided by $\omega^2$ is shown for increasing disorder
in Fig. \ref{triplet}.
Each panel shows $N(\omega)/\omega^2$ for one realization of disorder.
The low-energy end of the spectrum
(well below the van Hove singularity, and in the range which had
a linear dispersion in the absence of disorder) is well-fit by
\begin{equation}
N(\omega)=A\omega +B\omega^2,
\end{equation}
where $A$ and $B$ depend on both $\Delta/t$ and $Un/t$.

Fig. \ref{AB} shows the parameters $A$ and $B$ vs. disorder for
$Un/t=3.3$. (Qualitatively similar behavior is found for
$Un/t=5.0$ and $7.0$.)
For weak disorder the integrated density of states remains nearly
quadratic in $\omega$, consistent with the low energy excitations
being weakly perturbed phonons.
As disorder increases, however, a {\it linear} contribution to
$N(\omega)$ emerges, corresponding to a {\it constant} density
of states $g(\omega)$.
By the time the condensate is nearly completely depleted, the
linear contribution to $N(\omega)$ dominates.

\section{concluding remarks}

Gillis {\it et al.}\cite{Gillis} have measured the low
temperature (50 mK - 1 K) heat
capacity of thin $^4$He films adsorbed in porous Vycor glass.
At low coverages, they find that the heat capacity is linear,
with no
evidence of a superfluid transition.  This phase is identified
with the
insulating state of bosons localized by disorder, the ``Bose glass.''
Above a critical coverage (corresponding to several monolayers), the
low temperature phase is a superfluid, with a heat capacity that
varies as $T^2$.

Although the $^4$He is adsorbed as a few-monolayer film in the
Vycor, the pores are connected to form a three-dimensional network.
For sufficiently long wavelengths, three-dimensional behavior is
expected.  Why does the specific heat of the
superfluid vary as $T^2$ rather than
$T^3$, as expected for a three-dimensional condensate?
The explanation is that for
excitations with wavelengths less than the typical pore size
$\lambda_{\rm pore}$ (several hundred {\AA}ngstr\"{o}ms), the
connectivity of
the porous network is unimportant, and the density of states
for phonon-like
excitations will be that of a two-dimensional superfluid.
Thus above a crossover temperature $k_B T_x \sim h c /
\lambda_{\rm pore}$,
the specific heat should vary as $T^2$, until the roton contribution
becomes appreciable.  For an upper bound on $T_x$ we can use the
bulk speed of sound $c \sim 3 \times 10^4$ cm/sec, which gives a
crossover
temperature of 30 mK.  This is surely an overestimate, since at the
coverages studied by Gillis {\it et al.} the pores are not close to
being filled and the compressibility is therefore much less than
in bulk.
An alternative estimate using the speed of sound in
thin $^4$He films adsorbed on
graphite gives a crossover temperature of 1 mK.

How big a linear specific heat does one expect in the Bose glass?
If one assumes a constant density of bosonic excitations in the
Bose glass
(as we found for the strongly disordered superfluid), then the
observed linear
specific heat translates to roughly one mode per particle per
10 $\mu eV$.
This energy scale is comparable to the quantum confinement
energy of a
${}^4 {\rm He}$ atom trapped in a pore several
hundred {\AA}ngstr\"{o}ms in diameter.

We have presented numerical solutions of the disordered Bose Hubbard
model in the Bogoliubov approximation.  This approximation correctly
captures the long wavelength properties of the clean Bose condensate,
and is equivalent to the random phase approximation.  It represents an
expansion about the weakly disordered and weakly interacting limit,
with a small parameter given by the depletion of the condensate.
We find that weak disorder hardly affects the condensate fraction
or the superfluid
density.  Instead, the condensate distorts to screen the
imposed random potential. Interactions help stabilize the condensate,
and prevent its collapse into a macroscopically occupied
localized state.

For strong disorder the condensate fraction and superfluid density are
reduced, and ultimately vanish for sufficiently large disorder
(although
the Bogoliubov approximation is no longer controlled by this point).
Our calculation therefore cannot access the critical properties of
the superfluid-insulator transition.  The Bogoliubov calculation,
however, does suggest a promising variational state for the
Bose glass.

The clean Bose condensate has a linear low energy density of states
in two dimensions, which implies a low temperature specific heat that
varies as $T^2$, as observed. We find that with increasing
disorder a constant
density of states appears at low energy.  This constant density of
states dominates as the condensate fraction and superfluid density
become small, and leads to a linear low temperature specific heat.
With our small sample sizes, we could not determine the extent to
which these excitations are localized.

\acknowledgments
We thank D.P. Arovas, N. Trivedi, A. Garcia and P. Lammert
for useful discussions,
and  M.H.W. Chan for sending us a preprint of ref. \onlinecite{Gillis}
before publication.
K.G.S. was supported by a National Defense Science and Engineering
Fellowship.  D.S.R. was supported by the National Science
Foundation through PYI grant DMR-91-57414, and acknowledges a
grant from the Alfred P. Sloan Foundation.

\vfil\eject

\begin{figure}
\caption{
The condensate wavefunction $\phi_0$ is concentrated at
the minima of the ``bare'' potential $V(i)$.  The effective potential
felt by the bosons is therefore increased in the regions of
high condensate
density, so that the resulting self-consistently determined potential
$\tilde{V}(i)$ resembles the original potential $V(i)$ with its lowest
values lopped off. (Here $Un/t=3.3$ and $\Delta/t=10.0$)
\label{Vscreen-fig}}
\end{figure}

\begin{figure}
\caption{
The autocorrelation function of the screened potential.  Note
that the screened potential does not develop any long-range
correlations. (${\cal V}=210$ sites and $Un/t=3.3$)
\label{autocorrelation}}
\end{figure}

\begin{figure}
\caption{
An example of the participation ratio for the Hartree
quasiparticle states
$\phi_\lambda$ as a function of their energy $\epsilon_\lambda$
for several system sizes. ($Un/t=3.3$ and $\Delta/t=10.0$)
\label{Plambda-2D}}
\end{figure}

\begin{figure}
\caption{
The condensate fraction ${\cal N}_0/{\cal N}$ versus
disorder $\Delta/t$ for several interaction strengths,
extrapolated to
${\cal V} \rightarrow\infty$.
\label{depletion}}
\end{figure}

\begin{figure}
\caption{
The density $\langle G | n_i | G \rangle$ of the Bogoliubov
ground state  versus the density ${\cal N}|\phi_0(i)|^2$ of
the Hartree state.  Note that
despite the large depletion of the condensate in the Bogoliubov
state, the total density is well approximated by the density of the
completely condensed state. ($Un/t=3.3$)
\label{densities}}
\end{figure}

\begin{figure}
\caption{
The transverse component of the current-current response function
 for the $12 \times 13$ lattice. ($Un/t=3.3$)
\label{chi}}
\end{figure}

\begin{figure}
\caption{
${\cal N}_{\rm eff}/{\cal N}$ and ${\cal N}_n/{\cal N}$ vs.
disorder for the $12 \times 13$ lattice. ${\cal N}_s=
{\cal N}_{\rm eff}-{\cal N}_n$.  ($Un/t=3.3$)
\label{Neff}}
\end{figure}

\begin{figure}
\caption{
The superfluid fraction as obtained through the twist method for
several
interaction strengths, extrapolated to ${\cal V}\rightarrow \infty$.
\label{twist}}
\end{figure}

\begin{figure}
\caption{
$N(\omega)/\omega^2$ vs. $\omega$ for $Un/t=3.3$ and
(a)$\Delta/t=0.0$  (b)$\Delta/t=10.0$
(c)$\Delta/t=18.0$.  As disorder increases, $N(\omega)/\omega^2$
diverges, indicating a deviation from the form $N(\omega) \sim
\omega^d$. (${\cal V}=210$ sites.)
\label{triplet}}
\end{figure}

\begin{figure}
\caption{
Coefficients of the linear and quadratic parts of the integrated
density of states $N(\omega)=A\omega +B\omega^2$ vs. disorder
(210 site
system).  As disorder is increased,
the linear term develops, indicative of a glassy system.
\label{AB}}
\end{figure}


\begin{references}
\bibitem{inertness}{Although it is convenient to think of the ``inert
layer'' as perfectly
independent of the condensate that forms ``above'' it, this
picture is,
of course, an oversimplification.  Exchange and ordinary scattering
between the bosons in the ``inert layer'' and those in the
condensate allows
the nominally localized bosons to delocalize through virtual
excitation
into the extended condensate.}
\bibitem{HFexplain}{Note that while (\ref{Hartree}) is symmetric
under particle interchange, it is not the most general Hartree-Fock
state, which would be constructed by allowing each particle to occupy
its own single-particle state, and then explicitly symmetrizing.}
\end{references}
\end{document}